\documentclass[11pt]{article}
\usepackage{amssymb}

\usepackage{graphicx}
\usepackage{amsbsy}

\begin{document}

\title{\begin{flushright} \footnotesize{CECS-PHY-01/04}\\
\footnotesize{USACH-FM-01/09} \end{flushright} \bigskip Canonical approach
to 2D supersymmetric WZNW model coupled to supergravity}
\author{Olivera Mi\v {s}kovi\' c$^*$ and Branislav Sazdovi\' c$^\dagger$ \medskip \\
{\small \textit{$^*$Centro de Estudios Cient\'{\i }ficos, Casilla 1469,
Valdivia, Chile. }}\\
{\small \textit{$^* $Departamento de F\'{\i }sica, Universidad de Santiago
de Chile, Casilla 307, Santiago 2, Chile.}}\\
{\small \textit{$^*$Institute ``Vin\v {c}a'', Department for Theoretical
Physics, P.O.Box 522, Belgrade 11001, Yugoslavia.}}\\
{\small \textit{$^\dagger$Institute of Physics, P.O.Box 57, Belgrade 11001,
Yugoslavia.}}}
\maketitle

\begin{abstract}
Starting from the known representation of the Kac-Moody algebra in terms of
the coordinates and momenta, we extend it to the representation of the super
Kac-Moody and super Virasoro algebras. Then we use general canonical method
to construct an action invariant under local gauge symmetries, where
components of the super energy-momentum tensor $L_\pm$ and $G_\pm$ play the
role of the diffeomorphisms and supersymmetries generators respectively. We
obtain covariant extension of WZNW theory with respect to local
supersymmetry as well as explicit expressions for gauge transformations.%
\newline
\newline
PACS numbers: 04.60.Kz, 12.60.Jv and 04.20.Fy
\end{abstract}



\section{Introduction}

The two-dimensional Wess-Zumino-Novikov-Witten (WZNW) model \cite{wess} has
been discussed from various viewpoints. For example, at the infrared-stable
fixed point it provides non-Abelian bosonization rules describing
non-interacting massless fermions. It is conformally invariant and exactly
soluble theory.

The Kac-Moody (KM) and Virasoro algebras play a vital role in understanding
the model and for constructing its Lagrangian, particularly if it is coupled
to the gauge fields. Coupling to 2D gravity is straight because the
components of the energy-momentum tensors $L_\pm$, as reparametrization
generators, satisfy anomaly free Virasoro algebra. On the other hand, the
currents $j_{\pm a}$ satisfy the KM algebra with central charge, so it is
not possible to gauge the full non-Abelian symmetry group, but only the
``anomaly free'' subgroup. Both features are consequences of the fact that
WZNW model describes axial anomaly while the reparametrization symmetry is
not anomalous.

Supersymmetric extension of the WZNW model has also been studied \cite
{vecchia} using superfield formalism as well as the component notation. It
was shown that it was possible to choose the fermionic field such that
fermions were completely decoupled.

In this paper we are going to construct supersymmetric WZNW model coupled to
2D supergravity. In our approach the super Kac-Moody (SKM) and super
Virasoro algebras will play important role, because the super WZNW (SWZNW)
model will appear as a Lagrangian realization of the super Virasoro algebra.
We will use the Hamiltonian formalism where the generators, satisfying the
super Virasoro Poisson brackets (PB) algebra, are the constraints so that
the components of the metric tensor and Rarita-Schwinger field will appear
naturally as Lagrange multipliers.

In Sec. II we introduce general canonical formalism in order to construct a
gauge invariant effective action from the known representation of the
constraints in terms of coordinates and momenta. Then we extend the known
representation of the KM algebra to the representation of the SKM algebra.
With the help of these expressions we construct the representation of the
energy-momentum super tensor, firstly in terms of superfields and then in
component notation. The components of super energy-momentum tensor are
ordinary energy-momentum tensor and supersymmetric generator and they
satisfy the supersymmetric extension of Virasoro algebra which is a starting
point for our construction.

In Sec. III we construct the effective Lagrangian for SWZNW model in
external supergravity field. We treat the components of energy-momentum
super tensor as the constraints and use the general canonical formalism to
obtain the corresponding Lagrangian. These components are the first class
constraints enabling us to find the gauge symmetry transformations
(reparametrizations and supersymmetry transformations) of the matter and the
gauge multiplets.

The Lagrangian constructed in Sec. III is not manifestly Lorentz invariant
because, in the Hamiltonian formalism, only some components of the gauge
fields appear. In Sec. IV we introduce additional components to complete the
gauge multiplet, so that we obtain the full covariant description of the
fields and the full local gauge invariance. These new components are pure
gauge, because they can be gauge away by local Weyl, local Lorentz and local
super Weyl symmetries.

Sec. V is devoted to conclusion. In the Appendix we introduce the notation
and establish connection between our Hamiltonian components and
corresponding Lagrangian fields.


\section{Super Virasoro Generators}

We are going to use the canonical formalism \cite{henneaux} to construct an
action invariant under the gauge transformations for a given algebra of the
group $G$. Firstly, we need a representation of this algebra as the Poisson
brackets algebra
\begin{equation}  \label{general1}
\{ G_m,G_n\} ={U_{mn}}^rG_r\, ,\quad\{ G_m,H_0\} ={V_m}^rG_r\, ,
\end{equation}
which elements $G_m$ and the Hamiltonian $H_0$ are functions of the
coordinates $q^i$ and canonically conjugate momenta $p_i$. Note that, by
definition, $G_m$ are the first class constrains. Then the canonical action,
defined in usual way as
\begin{equation}  \label{action}
I[q,p,u]=\int dx\,(p_i \dot q^i-H_0-u^m G_m)\, ,
\end{equation}
is invariant under the gauge transformations of any quantity on phase space $%
F(q^i,p_i)$,
\begin{equation}  \label{general2}
\delta F=\{ F,\varepsilon^m G_m\}\, ,
\end{equation}
and of the Lagrange multipliers $u^m$,
\begin{equation}  \label{general3}
\delta u^m=\dot\varepsilon^m +u^r\varepsilon^s{U_{sr}}^m+ \varepsilon^r{V_r}%
^m\, .
\end{equation}
The multipliers will be identified as gauge fields, later.

Similar approach has been used for construction of the action for W-strings
propagating on group manifold and on curved backgrounds \cite{mikovic} and
for 2D induced gravity \cite{popovic}. Covariant extension of the WZNW model
with respect to arbitrary internal group has been obtained in \cite
{sazdovic1} and with respect to $SL(2,R)$ internal group and diffeomorphisms
in \cite{blagpop}, by the same method. In the last example, from explicit
expressions of the KM currents $j_{\pm a}$, the related group invariants $%
t_\pm=\mp\frac{1}{2k}\, \gamma^{ab} j_{\pm a}j_{\pm b}$ have been
constructed, representing components of the energy-momentum tensor and
satisfying two independent Virasoro algebras without central charges. Then
the covariant extension of the WZNW model has been obtained using the first
class constrains $t_\pm$ as generators of \textit{diffeomorphisms}. In the
present paper we are going to supersymmetrize above approach.

Following the same steps, let us firstly construct the representation of the
SKM algebra. We shall start from the known representation of KM algebra and
then introduce a fermionic field in such a way that extended algebra closes
and forms SKM.

Let us introduce notation. The field $g$ is a mapping from a two-dimensional
Riemannian spacetime $\Sigma$ to a semi-simple Lie group $G$, parametrized
by local coordinates $q^i$, $g=g(q^i)$. The generators of the group $G$, $%
t_a $, satisfy the Lie algebra $[t_a,t_b]={f_{ab}}^c\, t_c$. The expansions
of one-forms $g^{-1}dg\equiv dq^iE^a_{+i}\, t_a$ and $gdg^{-1}\equiv
dq^iE^a_{-i}\, t_a$ define vielbeins on the group manifold $E^a_{\pm i}$.
The Cartan metric in the tangent space is $\gamma_{ab}=\frac 12 \mathop%
\mathrm{Tr}\nolimits (t_a t_b)$ (the trace is taken in the adjoint
representation of $G$), and in the coordinate basis $\gamma_{ij}\equiv
E_{+i}^a\, E_{+j}^b\, \gamma_{ab}=E_{-i}^a\, E_{-j}^b\, \gamma_{ab}$.
Variables $E_{\pm a}^i$ and $\gamma^{ij}$ are inverses of the $E^a_{\pm i}$
and $\gamma_{ij}$ respectively. On the basis of the theorem that any closed
form is locally exact, the equation $d\left(gdg^{-1}gdg^{-1}gdg^{-1}%
\right)=0 $ can be written in the form $\frac 12 \mathop{\rm Tr}\nolimits
\left( gdg^{-1}\right)^3=-6d\tau$, where $\tau\equiv\frac
12\,dq^idq^j\tau_{ij}$ is a two-form. Rewriting in components, it becomes $%
\partial _i\tau_{jk}+\partial _j\tau_{ki}+\partial _k\tau_{ij}= \pm \frac{1}{%
2}\, f_{abc}E_{\mp i}^{a} E_{\mp j}^{b}E_{\mp k}^{c}$.

Now, we are ready to introduce the phase space representation of bosonic
part of KM currents in 2D Minkowski space-time $(\tau,\sigma)$ (well known
from \cite{sazdovic1}, \cite{sazdovic2}) as a
\begin{equation}  \label{current}
j_{\pm a}= -E^i_{\pm a}\,j_{\pm i}\, ,\qquad j_{\pm i}\equiv p_i+kP_{\pm
ij}\,{q^{\prime}}^j\, ,
\end{equation}
where the prime denotes the space derivative, and momentum independent part
is
\begin{equation}  \label{p}
P_{\pm ij}\equiv\tau_{ij}\pm\frac 12\,\gamma_{ij}\, .
\end{equation}
PB of currents (\ref{current}) defines two independent KM algebras of the
group $G$, with the central charges $\pm k$:
\begin{equation}  \label{KMalgebra}
\{ j_{\pm a}(x),j_{\pm b}(y)\}={f_{ab}}^c\, j_{\pm c}(x)\, \delta(\sigma_x
-\sigma_y)\pm k\,\gamma_{ab}\delta^{\prime}(\sigma_x-\sigma_y)\, .
\end{equation}

In order to supersymmetrize the above algebra, let us introduce a Lie
algebra valued fermionic fields $\hat\chi_{\pm a}$. Because the fermionic
part of the Lagrangian should be linear in time derivative, there always
exist the second class constrains $S_{\pm a}\equiv\pi_{\pm
a}-ik\,\hat\chi_{\pm a}$ linear in coordinate $\hat\chi_{\pm a}$ and in
corresponding canonical momenta $\pi_{\pm a}$. Dirac brackets for the
fermionic fields are $\{\hat\chi_{\pm a},\hat\chi_{\pm b}\}^*=-\frac{i}{2k}%
\, \gamma_{ab}\,\delta$, while for bosonic currents $j_{\pm a}$ they remain
the same as the PB. So, we can start from the relation (\ref{KMalgebra}) and
\begin{equation}  \label{db}
\{\hat\chi_{\pm a}(x),\hat\chi_{\pm b}(y)\} =-\frac{i}{2k}\, \gamma_{ab}\,
\delta(\sigma_x-\sigma_y)\, ,
\end{equation}
where, from now, we omit star because of simplicity. Note that in both
bosonic and fermionic cases all quantities of the opposite chirality commute.

It is easy to check that bilinears in the fermionic fields $%
\widetilde\jmath_{\pm a}\equiv-ik\,f_{abc}\,\hat\chi^b_\pm \hat\chi^c_\pm$
satisfy the KM algebra without central charges
\begin{equation}  \label{auxilliar1}
\{ \widetilde\jmath_{\pm a}(x),\widetilde\jmath_{\pm b}(y)\} = {f_{ab}}^c\,
\widetilde\jmath_{\pm c}(x)\delta(\sigma_x-\sigma_y)\, ,
\end{equation}
and have nontrivial brackets with $\hat\chi_{\pm b}$,
\begin{equation}  \label{auxilliar2}
\{\widetilde\jmath_{\pm a}(x),\hat\chi_{\pm b}(y)\} = {f_{ab}}^c\,
\hat\chi_{\pm c}(x)\delta(\sigma_x-\sigma_y)\, .
\end{equation}
We can introduce new currents $J_{\pm a}\equiv j_{\pm
a}+\widetilde\jmath_{\pm a}$ (such that KM algebra remains unchanged) which
with its supersymmetric partners $\hat\chi_{\pm a}$ satisfy two independent
SKM algebras:
\begin{eqnarray}  \label{sKMalgebra1}
\{ J_{\pm a}(x),J_{\pm b}(y)\} &=& {f_{ab}}^c\, J_{\pm c}(x)
\delta(\sigma_x-\sigma_y)\pm k\,\gamma_{ab}\,\delta^{\prime}(\sigma_x-
\sigma_y)\, ,  \nonumber \\
\{ J_{\pm a}(x),\hat\chi_{\pm b}(y)\} &=& {f_{ab}}^c\, \hat\chi_{\pm c}(x)
\delta(\sigma_x-\sigma_y)\, ,  \nonumber \\
\{ \hat\chi_{\pm a}(x),\hat\chi_{\pm b}(y)\} &=&-\frac{i}{2k}\,\gamma_{ab}\,
\delta(\sigma_x-\sigma_y)\, .
\end{eqnarray}

Next step is the construction of the components of energy-momentum super
tensors as functions of the SKM currents, keeping in mind that they have to
be group invariants. The easiest way to do this is to introduce the
superfields

\begin{equation}  \label{Scurrent}
I_{\pm a}(z)\equiv\sqrt 2k\hat\chi_{\pm a}(x)+\theta_\mp J_{\pm a}(x) \, ,
\end{equation}
and rewrite the algebra (\ref{sKMalgebra1}) in the form
\begin{equation}  \label{sKMalgebra2}
\{ I_{\pm a}(z_1),I_{\pm b}(z_2)\}=\delta_{\pm 12}{f_{ab}}^c\, I_{\pm
c}(z_1)-ik\,\gamma_{ab}\,D^\pm\delta_{\pm 12}\, ,
\end{equation}
where $D^\pm\equiv\frac{\partial} {\partial\theta_{\mp}}\,\mp i\theta_\mp%
\frac{\partial}{\partial\sigma}$ is the super covariant derivative, while $%
\delta_{\pm 12}= (\theta_{\pm 1}-\theta_{\pm 2})\delta(\sigma_1-\sigma_2)$
is a generalization of the Dirac $\delta$-function to the super $\delta$%
-function. Derivative is always taken over the first argument of $\delta$%
-function. Notation is given in the Appendix.

Up to the third power of $I_{\pm a}$, there are only two invariants $%
\gamma_{ab}\,D^\pm I^a_\pm I^b_\pm$ and $f_{abc}\,I^a_\pm I^b_\pm I^c_\pm$.
Note that $\gamma_{ab}\,I^a_\pm I^b_\pm$ is identically equal to zero
because super currents are odd variables. A requirement for the closed
algebra determines the ratio of coefficients multiplying two invariants, so
we take for super energy-momentum tensor
\begin{equation}  \label{Stei}
T_\pm\equiv\mp\frac{1}{2k}\,\left(\gamma_{ab}\,D^\pm I^a_\pm I^b_\pm+ \frac{i%
}{3k}\,f_{abc}\,I^a_\pm I^b_\pm I^c_\pm\right)\, .
\end{equation}
In components notation, we have
\begin{equation}  \label{Ctei}
T_\pm=\mp G_\pm +\theta_\mp L_\pm\, ,
\end{equation}
where the bosonic part is
\begin{equation}  \label{components1}
L_\pm=\mp\frac{1}{2k}\left(J^a_\pm J_{\pm a}\mp 2ik^2\hat\chi^{{\prime}
a}_\pm \hat\chi_{\pm a}+2ikf_{abc}\,\hat\chi^a_\pm\hat\chi^b_\pm
J^c_\pm\right)\, ,
\end{equation}
while its supersymmetric partner is
\begin{equation}  \label{components2}
G_\pm=\frac{1}{\sqrt 2}J^a_\pm \hat\chi_{\pm a}+\frac{i\sqrt 2 k}{3}
f_{abc}\,\hat\chi^a_\pm\hat\chi^b_\pm \hat\chi^c_\pm\, .
\end{equation}
It is useful to express (\ref{components1}) and (\ref{components2}) in terms
of quantities $j_{\pm a}$ and $\hat\chi _{\pm a}$:
\begin{eqnarray}  \label{components3}
L_\pm&=&\mp\frac{1}{2k}j^a_\pm j_{\pm a}+ ik\hat\chi^{\prime a}_\pm
\hat\chi_{\pm a}\, ,  \nonumber \\
G_\pm&=&\frac{1}{\sqrt 2}\,j^a_\pm \hat\chi_{\pm a}-\frac{ik} {3\sqrt 2}
f_{abc}\,\hat\chi^a_\pm\hat\chi^b_\pm \hat\chi^c_\pm\,.
\end{eqnarray}
Using the SKM algebra (\ref{sKMalgebra1}), we can obtain the following
brackets between the components of energy-momentum tensor and currents
\begin{eqnarray}  \label{Apb}
\{L_\pm,J_{\pm a}\}=-J_{\pm a}\,\delta^{\prime}\, ,\qquad~\quad\quad &&
\{G_\pm,J_{\pm a}\}= \pm\frac{k}{\sqrt 2}\,\hat\chi_{\pm
a}\,\delta^{\prime}\, ,  \nonumber \\
\{L_\pm,\hat\chi_{\pm a}\}= \frac 12 \, (\hat\chi^{\prime}_{\pm a}\,\delta -
\hat\chi_{\pm a}\delta^{\prime})\, ,&& \{G_\pm,\hat\chi_{\pm a}\} =-\frac{i}{%
2\sqrt 2k}\,J_{\pm a}\,\delta\, ,
\end{eqnarray}
as well as the brackets between the components of energy-momentum tensor
themselves
\begin{eqnarray}  \label{virasoro1}
\{ L_\pm,L_\pm\}&=& -(L^{\prime}_\pm\delta+2L_\pm\delta^{\prime})~=~-\left[
L_\pm(x)+ L_\pm(y)\right] \delta^{\prime}\, ,  \nonumber \\
\{ G_\pm,G_\pm\}&=& \pm\frac i2 L_\pm\delta \, ,  \nonumber \\
\{ L_\pm,G_\pm\}&=& -\frac 12 \left(
G^{\prime}_\pm\delta+3G_\pm\delta^{\prime}\right) \, ,  \nonumber \\
\{ G_\pm,L_\pm\}&=& -\frac 12 \left(
2G^{\prime}_\pm\delta+3G_\pm\delta^{\prime}\right)\, .
\end{eqnarray}
In terms of the super fields, we have instead of (\ref{Apb})
\begin{equation}  \label{TI}
\{ T_\pm(z_1),I_{\pm a}(z_2)\}=\pm\frac i2 \, \left(D^\pm I_{\pm a}D^\pm
\delta_{\pm 12}+I_{\pm a}{D^\pm}^2 \delta_{\pm 12} \right) \, ,
\end{equation}
and instead of (\ref{virasoro1})
\begin{equation}  \label{virasoro2}
\{ T_\pm(z_1),T_\pm(z_2)\}=\pm\frac i2\,\left( 2{D^\pm}^2 T_\pm
\,\delta_{\pm 12}+ D^\pm T_\pm\, D^\pm\delta_{\pm 12}+3T_\pm\,{D^\pm}%
^2\delta_{\pm 12}\right)\, .
\end{equation}
This is a supersymmetric extension of the Virasoro algebra \textit{without
central charge}. Since $L_\pm$ and $G_\pm$ are the first class constrains,
we shall apply the general canonical method to construct a theory invariant
under \textit{diffeomorphisms} generated by $L_\pm$ and under local \textit{%
supersymmetry} generated by $G_\pm$. Because we know from \cite{blagpop}
that, in the bosonic case with the similar approach, we have got covariant
extension of the WZNW theory with respect to diffeomorphisms, here we expect
to obtain covariant extension of WZNW theory with respect to local
supersymmetry.


\section{Effective Lagrangian and Gauge Transformations}

In order to construct a covariant theory we start with the generators
\begin{equation}  \label{case}
H_0=0\, , \qquad G_m=(L_-,L_+,G_-,G_+)\, ,
\end{equation}
with the explicit expressions given in equations (\ref{components3}), and
the PB algebra (\ref{virasoro1}) instead of the first equation (\ref
{general1}). According to (\ref{action}), we introduce the canonical
Lagrangian
\begin{equation}  \label{Clag}
\widehat{\mathcal{L}}=\dot q^ip_i +ik\,\dot{\hat\chi}^a_+\hat\chi_{+a}+ik\,
\dot{\hat\chi}^a_-\hat\chi_{-a}-h^-L_--h^+L_+-i\psi^-G_--i\psi^+G_+
\end{equation}
with multipliers $u^m=(h^-,h^+,\psi^-,\psi^+)$. Note that, on Dirac
brackets, the second class constrains are equal to zero $(S_{\pm a}=0)$, so
we have $\pi_{\pm a}= ik\,\hat\chi_{\pm a}$. We can eliminate remain
momentum variables with the help of their equations of motion
\begin{equation}  \label{momentum}
p^i=\frac{k}{h^--h^+}\left[\dot q^i+\left( h^+{{P_+}^i}_j- h^-{{P_-}^i}%
_j\right){q^{\prime}}^j +\frac{i}{\sqrt 2}\,\left(\psi^-\hat\chi^i_-+
\psi^+\hat\chi^i_+\right)\right]\, ,
\end{equation}
where $\hat\chi^i_\pm=E^i_{\pm a}\hat\chi^a_\pm$. On the equations of motion
(\ref{momentum}), the currents (\ref{current}) become
\begin{equation}  \label{j.motion}
j^i_\pm=\frac k2 \left[\hat\partial_\pm q^i +\frac{i}{\sqrt 2 \sqrt{-\hat g}}%
\,\left(\psi^-\hat\chi^i_- +\psi^+\hat\chi^i_+\right)\right]\, ,
\end{equation}
so that the Lagrangian (\ref{Clag}) can be written in the form:
\begin{eqnarray}  \label{lagrangian}
\widehat{\mathcal{L}}&=&\widehat{\mathcal{L}}_{WZ}+\widehat{\mathcal{L}}_f+%
\widehat{\mathcal{L}}_{int}  \nonumber \\
\widehat{\mathcal{L}}_{WZ}&=&-\frac k2 \sqrt{-\hat g}\,P_{-ij}\,
\hat\partial_-q^i\hat\partial_+q^j  \nonumber \\
\widehat{\mathcal{L}}_f&=&-ik\sqrt{-\hat g}\,\left( \hat\chi^a_+ \hat
D_-\hat\chi_{+a}+ \hat\chi^a_-\hat D_+\hat\chi_{-a}\right)  \nonumber \\
\widehat{\mathcal{L}}_{int}&=&\frac{ik}{2\sqrt 2}\,\left(
\psi^+\hat\chi_{+i}\hat\partial_+q^i+
\psi^-\hat\chi_{-i}\hat\partial_-q^i\right)\, .
\end{eqnarray}
Here, $\widehat{\mathcal{L}}_{WZ}$ and $\widehat{\mathcal{L}}_{f}$ are WZNW
and fermion Lagrangians respectively, covariantized in external super
gravitational fields $\hat g_{\mu \nu}$ and $\psi^\pm$, while $\widehat{%
\mathcal{L}}_{int}$ describes interaction between bosonic fields $q^i$ and
fermionic fields $\hat\chi_{\pm i}$. The tensor $\hat g_{\mu\nu}$ is
introduced instead of variables $(h^+,h^-)$ (see the Appendix)
\begin{equation}
\hat g_{\mu\nu}\equiv-\frac 12\, \left(
\begin{array}{cc}
-2h^+h^- & h^++h^- \\
h^++h^- & -2
\end{array}
\right)\, .
\end{equation}
Covariant derivatives, acting on fermionic fields $\hat\chi^a_\pm$, are
defined by
\begin{equation}  \label{act}
\hat D_\mp\hat\chi^a_\pm\equiv\hat\partial_\mp\hat\chi^a_\pm+\frac{i}{3\sqrt
2\sqrt{-\hat g}}\, {f^a}_{bc}\psi^\pm\hat\chi^b_\pm\hat\chi^c_\pm \pm\frac{i%
}{8\hat g} \,\psi^-\psi^+\,\hat\chi^a_\mp\, ,
\end{equation}
where $\hat\partial_\pm = {{\hat e}^\mu}_{\;\;\pm}\partial_\mu $, and ${{%
\hat e}^\mu}_{\;\;\pm}$ are also given in the Appendix.

The general canonical method provides a mechanism to write out gauge
symmetries of the Lagrangian (\ref{lagrangian}). Instead of relations (\ref
{general2}), with the help of (\ref{components3}), we find the following
gauge transformations of the fields,
\begin{eqnarray}  \label{fields}
\delta q^i&=&\frac 1k\, (\varepsilon^-j^i_- -\varepsilon^+j^i_+)- \frac{i}{%
\sqrt 2}\, (\eta^+\hat\chi^i_++\eta^-\hat\chi^i_-)\, ,  \nonumber \\
\delta\hat\chi^a_\pm&=&-\varepsilon^\pm\partial_1\hat\chi^a_\pm-\frac 12\,
\left(\partial_1\varepsilon^\pm\right) \hat\chi^a_\pm-\frac{1}{2k\sqrt 2}%
\,\eta^\pm\, J^a_\pm\, ,
\end{eqnarray}
and instead of (\ref{general3}) using (\ref{virasoro1}), we obtain the gauge
transformations of the multipliers,
\begin{eqnarray}  \label{multipliers}
\delta h^\pm&=&\partial_0\varepsilon^\pm+h^\pm\partial_1\varepsilon^\pm-
\varepsilon^\pm\partial_1 h^\pm \pm\frac i2\, \psi^\pm\eta^\pm\, ,  \nonumber
\\
\delta\psi^\pm&=& \frac 12 \, \psi^\pm\partial_1 \varepsilon^{\pm}
-\left(\partial_1\psi^\pm\right)\varepsilon^\pm+
\partial_0\eta^\pm+h^\pm\partial_1\eta^\pm-\frac 12\,
\left(\partial_1h^\pm\right)\eta^\pm\, .
\end{eqnarray}
Bosonic fields $\varepsilon^\pm$ and fermionic fields $\eta^\pm$ are
parameters of diffeomorphisms and local supersymmetry transformations
respectively.


\section{Lagrangian Formulation}

It turns out that the Lagrangian (\ref{lagrangian}) is invariant under the
following rescaling of fields by two arbitrary parameters $F(x)$ and $f(x)$:
\begin{eqnarray}  \label{rescaling1}
{{\hat e}^\pm}_{\;\;\mu}&\to&{e^\pm}_\mu\equiv e^{F\pm f}\, {{\hat e}^\pm}%
_{\;\;\mu}\, ,  \nonumber \\
\psi^\pm&\to&\psi_{\mp(\mp)}\equiv\frac{1}{2\sqrt{-\hat g}}\, e^{-\frac
12\,(F\mp 3f)}\,\psi^\pm \, ,  \nonumber \\
\hat\chi^a_\pm&\to&\chi^a_\pm\equiv e^{-\frac 12\,(F\pm
f)}\,\hat\chi^a_\pm\, .
\end{eqnarray}
As a consequence, we have:
\begin{eqnarray}  \label{rescaling2}
\sqrt{-\hat g}&\to&\sqrt{-g}=e^{2F}\,\sqrt{-\hat g}\, ,  \nonumber \\
\hat\partial_\pm&\to&\partial_\pm\equiv e^{-(F\pm f)}\,\hat\partial_\pm\,.
\end{eqnarray}
In terms of rescaled fields, the rescaled Lagrangian has the same form as
the original one (\ref{lagrangian}),
\begin{eqnarray}  \label{newL}
\mathcal{L}&=&\mathcal{L}_{WZ}+\mathcal{L}_f+\mathcal{L}_{int}\, ,  \nonumber
\\
\mathcal{L}_{WZ}&=&-\frac k2 \sqrt{-g}\,P_{-ij}\, \partial_-q^i\partial_+q^j
\, ,  \nonumber \\
\mathcal{L}_f&=&-ik\sqrt{-g}\,\left(\chi^a_+ D_-\chi_{+a}+
\chi^a_-D_+\chi_{-a}\right)\, ,  \nonumber \\
\mathcal{L}_{int}&=&\frac{ik}{\sqrt 2}\,\sqrt{-g}\,\left[
\psi_{-(-)}\chi_{+i}\partial_+q^i+
\psi_{+(+)}\chi_{-i}\partial_-q^i\right]\, ,
\end{eqnarray}
where $D_\mp\chi^a_\pm\equiv\partial_\mp\chi^a_\pm+\frac{i\sqrt 2}{3}\, {f^a}%
_{bc}\psi_{\mp(\mp)}\chi^b_\pm\chi^c_\pm\mp\frac i2\,\psi_{+(+)}
\psi_{-(-)}\,\chi^a_\mp$. Note that the term with derivatives over $F$ and $%
f $ vanishes because of nilpotency of the field $\hat\chi^{i}_\pm$.

Introduction of the new fields $F$ and $f$ gives additional gauge freedom to
the Lagrangian (\ref{newL}). It becomes invariant under the \textit{local
Weyl transformations}
\begin{eqnarray}  \label{weyl}
\delta_\sigma {e^\pm}_\mu&=&\sigma {e^\pm}_\mu\, ,  \nonumber \\
\delta_\sigma\psi_{\pm(\pm)}&=&-\frac 12\,\sigma\psi_{\pm(\pm)}\, ,
\nonumber \\
\delta_\sigma\chi^a_\pm&=&-\frac 12\,\sigma\chi^a_\pm \, ,
\end{eqnarray}
as a consequence of the transformation $\delta_\sigma F=\sigma$, while all $%
F $ independent fields remain Weyl invariant. Furthermore, the Lagrangian (%
\ref{newL}) does not change under the \textit{local Lorentz transformations}
\begin{eqnarray}  \label{lorentz}
\delta_\ell {e^\pm}_\mu&=&\mp\ell {e^\pm}_\mu\, ,  \nonumber \\
\delta_\ell \psi_{\pm(\pm)} &=& \pm\frac 32\,\ell\,\psi_{\pm(\pm)}\, ,
\nonumber \\
\delta_\ell \chi^a_\pm &=& \pm\frac 12\,\ell\,\chi^a_\pm\, ,
\end{eqnarray}
generated by the transformation $\delta_\ell f=-\ell$. The vielbein ${e^\pm}%
_\mu$ is a Lorentz vector, the fields $\psi_{\pm(\pm)}$ and $\chi^a_\pm$
transform like components of a spinor field with the spin $\frac 32$ and $%
\frac 12$ respectively, while all $f$ independent fields are Lorentz scalars.

The Lagrangian (\ref{newL}) depends only on two components $\psi_{\pm(\pm)}$
of the Rarita-Schwinger spinor field $\psi_{\mu\alpha}={e^a}%
_\mu\psi_{a(\alpha)}\,(\mu=0,1,\, \alpha=+,-)$. It means that we also have
the additional \textit{local super Weyl symmetry}
\begin{equation}  \label{Sweyl}
\delta_\lambda\psi_{\pm(\mp)} = \pm\lambda_\pm\, ,
\end{equation}
while all other fields are super Weyl invariant. Transformations (\ref{weyl}%
) -- (\ref{Sweyl}) can be written in a covariant form,
\begin{equation}  \label{all}
\left.
\begin{array}{rlll}
\mbox{Lorentz:} & \delta_\ell{e^a}_\mu =-\ell\,{\varepsilon^a}_b\,{e^b}%
_\mu\, , & \delta_\ell\psi_\mu =\frac 12 \,\ell\,\gamma_5\,\psi_\mu\, , &
\delta_\ell\chi^a=\frac 12\,\ell\gamma_5\,\chi^a\, , \medskip \\
\mbox{Weyl:} & \delta_\sigma{e^a}_\mu = \sigma{e^a}_\mu\, , &
\delta_\sigma\psi_\mu=\frac 12 \sigma\psi_\mu\, , & \delta_\sigma\chi^a=%
\frac 12\,\sigma\chi^a\, , \medskip \\
\mbox{super Weyl:} & \delta_\lambda{e^a}_\mu = 0\, , & \delta_\lambda\psi_%
\mu = \gamma_\mu\lambda\, , & \delta_\lambda\chi^a=0,
\end{array}
\right.
\end{equation}
where $\gamma_\mu\equiv{e^a}_\mu\gamma_a$. The representation of the $\gamma$%
-matrices is given in the Appendix. Note that the fields ${{\hat e}^\pm}%
_{\;\;\mu}$, $\psi^\pm$ and $\hat\chi^a_\pm$, introduced in the Hamiltonian
approach are Lorentz, Weyl and super Weyl invariants.

The fields $F$ and $f$ do not enter the original Lagrangian (\ref{lagrangian}%
) but are introduced by rescaling the fields (\ref{rescaling1}). Thus we
cannot find their change under diffeomorphisms and SUSY transformations just
applying the general Hamiltonian rules (\ref{general2}), (\ref{general3}).
We have introduced them in such a way that the new fields ${e^a}_\mu$ and $%
\psi_{\mu\alpha}$ have proper Lorentz, Weyl and super Weyl transformations.
Now, we demand that ${e^a}_\mu$ transforms like a vector and $%
\psi_{\mu\alpha}$ transforms like a Rarita-Schwinger field under general
coordinate transformations with a local parameter $\varepsilon^\mu (x)$ and
under $N=1$ supersymmetric transformations with a local spinor parameter $%
\zeta_\alpha (x)$. It means that (see for example \cite{hoker})
\begin{eqnarray}  \label{diffSUSY1}
\delta {e^a}_\mu&=&-\varepsilon^\nu\partial_\nu{e^a}_\mu-{e^a}_\nu
\partial_\mu\varepsilon^\nu -\frac i2 \,\bar\zeta\gamma^a\psi_\mu\, ,
\nonumber \\
\delta\psi_\mu&=&-\varepsilon^\nu\partial_\nu\psi_\mu-\psi_\nu\,
\partial_\mu\varepsilon^\nu+\frac 12\, \triangledown_\mu\zeta\, ,
\end{eqnarray}
where $\triangledown_\mu\zeta={e^a}_\mu\triangledown_a\zeta$ and $%
\triangledown_a\zeta= \left(\partial_a+\frac 12\gamma_5\omega_a\right)\,
\zeta$. Writing out in components, it gives
\begin{eqnarray}  \label{diffSUSY2}
\delta(F\pm f) &=&h^\pm\partial_1\varepsilon^0-\partial_1\varepsilon^1-
\varepsilon^\mu\partial_\mu (F\pm f) \pm ie^{- (F\pm f)} \zeta_\mp\psi_{1
\mp} \, ,  \nonumber \\
\delta h^\pm&=&
\partial_0\varepsilon^1-h^\pm(\partial_0\varepsilon^0-\partial_1%
\varepsilon^1)- (h^\pm)^2 \partial_1\varepsilon^0-
\varepsilon^\mu\partial_\mu h^\pm \mp \frac i2\,e^{-\frac 12 (F\pm f)}
\zeta_\mp\psi^\pm\, ,  \nonumber \\
\delta\psi^\pm&=& \left[\frac 12\, \partial_1(\varepsilon^1-h^\pm
\varepsilon^0) +\frac
12\,(\partial_1h^\pm)\varepsilon^0-\partial_0\varepsilon^0 -h^\pm
\partial_1\varepsilon^0\right]\psi^\pm-\varepsilon^\mu\partial_\mu\psi^\pm
\nonumber \\
&&+\left(\partial_0+h^\pm\partial_1-\frac 12\, \partial_1h^\pm
\right)\left(\zeta_\mp e^{-\frac 12\,(F\pm f)}\right)+ \frac
i4\,\zeta_\mp\psi_{\pm(\mp)}\psi^\pm\, ,
\end{eqnarray}
where $\psi^\mp\equiv 2e^{-\frac 12 \, (F\mp f)}\left(
\psi_{0\pm}+h^\mp\psi_{1\pm}\right)$ in according with (\ref{rescaling1}).

In order to establish relation between Hamiltonian and Lagrangian
transformations we should compare the Hamiltonian transformations (\ref
{multipliers}) with Lagrangian one, the last two equations of (\ref
{diffSUSY2}). We find that we can identify them choosing the following
relation between the gauge parameters
\begin{eqnarray}  \label{param}
\varepsilon^\pm\equiv \varepsilon^1-h^\pm\varepsilon^0\, ,\qquad
\eta^\pm\equiv 2 \zeta_\mp e^{-\frac 12\,(F\pm f)}-\varepsilon^0 \psi^\pm\, ,
\end{eqnarray}
and imposing gauge fixing $\psi_{\pm(\mp)}=0$, because in the Hamiltonian
approach all quantities are super Weyl invariant. Substituting equations of
motion (\ref{j.motion}) in (\ref{fields}) we can obtain the momentum
independent formulation of transformation law of matter variables $q^i$, $%
\hat\chi^a_\pm$. In terms of the Lagrangian variables $\varepsilon^\mu$ and $%
\zeta_\pm$, we have

\begin{eqnarray}  \label{density}
\delta q^i&=&-\varepsilon^\mu\,\partial_\mu q^i- i\sqrt 2\, \left(\zeta_-
\chi^i_++\zeta_+\chi^i_-\right) \, ,  \nonumber \\
\delta\hat\chi^a_\pm&=&-\varepsilon^\mu\,\partial_\mu \hat\chi_\pm ^a
+\varepsilon^0 \sqrt{-\hat g}\hat\triangledown_\mp \hat\chi_\pm ^a + \frac
12\, \left(h^\pm \partial_1 \varepsilon^0 - \partial_1 \varepsilon^1
\right)\hat\chi_\pm ^a+  \nonumber \\
&+& \frac{1}{2k\sqrt 2}\, \left( \varepsilon^0\psi^\pm -2 \zeta_\mp e^{-%
\frac{1}{2} (F\pm f)} \right) J^a_\pm\, .  \nonumber \\
\end{eqnarray}

In the flat space limit ${e^a}_\mu\to\delta^a_\mu$ and $\psi_\mu\to 0$, the
Lagrangian (\ref{newL}) becomes
\begin{equation}  \label{Flagrangian}
\mathcal{L}_0=-\frac k2 \,P_{-ij}\,
\partial_-q^i\partial_+q^j-ik\left(\chi^a_+\partial_-\chi_{+a}+
\chi^a_-\partial_+\chi_{-a}\right)\, ,
\end{equation}
and bosonic and fermionic parts are decupled. The first term is the bosonic
Wess-Zumino action, while the second one is the Lagrangian of free spinor
fields $\chi^a_\pm$. The Lagrangian (\ref{Flagrangian}) describes the $N=1$
supersymmetric WZNW theory \cite{vecchia}.


\section{Conclusion}

Using the general canonical method we construct the Lagrangian for the SWZNW
model coupled to 2D supergravity. The basic ingredients of our approach are
the symmetry generators, which are functions of the coordinates and momenta
and satisfy the SKM and super Virasoro PB algebras. Application of the
Hamiltonian method naturally incorporates gauge fields (components of the
metric tensor and Rarita-Schwinger fields) as Lagrange multipliers of the
symmetry generators. This method also gives a prescription for finding gauge
transformations for both the matter and gauge fields.

The result of Sec. II is the representation of the super Virasoro algebra
elements (the components of energy-momentum tensor $L_\pm$ and
supersymmetric generators $G_\pm$) as functions of the coordinates and
momenta, eq. (\ref{components3}). In this approach, firstly we had to
construct the KM and SKM algebras, where we introduced superfields as a
useful method for going from the SKM algebra to the super Virasoro one.

In Sec. III the effective Lagrangian for the theory which symmetry is
defined by super Virasoro algebra (\ref{virasoro1}) has been constructed.
Treating generators of the algebra as the first class constraints, the
general canonical method (introduced in Sec. II) has been applied. After
elimination of the bosonic momentum variables on the basis of their
equations of motion, we obtain the effective action containing WZNW theory,
fermionic Lagrangian and the part describing their interaction in external
supergravity field. The Hamiltonian version of reparametrization and
supersymmetry transformations has been found for the matter fields $q^i$ and
$\hat\chi^a_\pm$ (\ref{fields}), and for the Lagrange multipliers $h^\pm$
and $\psi^\pm$ (\ref{multipliers}).

The Hamiltonian formalism deals with the Lagrangian multipliers $h^\pm$ and $%
\psi^\pm$, which are just the part of the gauge fields necessary to
represent the symmetry of the algebra. In Lagrangian formulation we need the
covariant description of the fields. In order to complete vielbeins $e^a_\mu$%
, it was necessary to introduce the new bosonic components $F$ and $f$,
while for completing the Rarita-Schwingerfields $\psi_{\mu \alpha}$ we need
the new fermionic fields $\psi_{\pm (\mp)}$ . The new components are not
physical because they do not appear in the Lagrangian, but they give
additional gauge freedom corresponding to the additional gauge symmetries.
These are local Weyl and local Lorentz symmetries for the bosonic fields $F$
and $f$ respectively, and local super Weyl symmetry for the fermionic field $%
\psi_{\mp(\pm)}$ . The fields $F$ and $f$ are not parts of the Hamiltonian
formalism, so we find their transformation laws under reparametrizations and
supersymmetry requiring that vielbeins ${e^a}_\mu$ transform as vectors and $%
\psi_{\mu \alpha}$ transform as a Rarita-Schwinger field. Consequently, in
Sec. IV we establish the complete relation between Hamiltonian and
Lagrangian formulations. We find the connection between corresponding
fields, gauge parameters and gauge transformations.

We can conclude that our main result is derivation of the Lagrangian for
SWZNW model coupled to supergravity, eq. (\ref{newL}). By construction, this
Lagrangian is invariant under local Lorentz, local Weyl and local super Weyl
transformations (\ref{all}), and under reparametrizations and N=1
supersymmetry transformations, (\ref{diffSUSY1}) and (\ref{density}).

Thanks to using the canonical approach, we can interpret the results in a
different way, as the complete canonical analysis of our Lagrangian. It
means that momenta are defined by eq. (\ref{momentum}), the canonical
Hamiltonian is zero (see (\ref{case})) and $L_\pm$ and $G_\pm$ defined in (%
\ref{components3}) are the first class constraints satisfying super Virasoro
PB algebra (\ref{virasoro1}).

In the flat space limit we reproduce the result of ref. [2].

\appendix

\section{Notation}

At each point of the curved two-dimensional space-time $\Sigma$ with
signature $(-\,+)$ and coordinates $x^\mu$ $(\mu =0,1)$, there is a \textit{%
light cone} basis of 1-forms $dx^a\equiv {e^a}_\mu dx^\mu$. Vielbeins ${e^a}%
_\mu$ are expressed in terms of variables $(h^-,h^+,F,f)$ as
$$
\label{basis} {e^\pm}_\mu=e^{F\pm f}{{\hat e}^\pm}_{\;\;\mu}\, ,\quad {{\hat
e}^a}_{\;\;\mu} =\frac 12\, \left(
\begin{array}{cc}
-h^+ & 1 \\
h^- & -1
\end{array}
\right)\quad (a=+,-;~ ~ \mu=0,1)\, . \eqno (A.1)
$$
Inverse vielbeins ${e^\mu}_a$ (${e_a}^\mu {e_\mu}^b=\delta^b_a$ and ${e_\mu}%
^a {e_a}^\nu=\delta^\nu_\mu$) are
$$
\label{tangent} {e^\mu}_\pm=e^{-(F\pm f)}{{\hat e}^\mu}_{\;\;\pm}\, ,\quad {{%
\hat e}^\mu}_{\;\;a} =\frac{2}{h^--h^+}\, \left(
\begin{array}{cc}
1 & 1 \\
h^- & h^+
\end{array}
\right) \, . \eqno (A.2)
$$
Related basis of tangent vectors $\partial_a \equiv {e^\mu}_a\partial_\mu$
can be written as
$$
\label{tang.expli} \partial_\pm=e^{-(F\pm f)}\hat\partial_\pm\, ,\quad
\hat\partial_\pm=\frac{2}{h^--h^+}\, (\partial_0+h^\mp\partial_1)\, . \eqno %
(A.3)
$$
It follows from $(A.1)$ that the components of the metric tensor $%
g_{\mu\nu}=\eta_{ab}\,{e^a}_\mu {e^b}_\nu$ are
$$
\label{metric} g_{\mu\nu}=e^{2F}\hat g_{\mu\nu}\, ,\qquad \hat
g_{\mu\nu}\equiv-\frac 12\, \left(
\begin{array}{cc}
-2h^+h^- & h^++h^- \\
h^++h^- & -2
\end{array}
\right)\, , \eqno (A.4)
$$
while the inverse metric $g^{\mu\nu}$ is
$$
\label{inv.metric} g^{\mu\nu}=e^{-2F}\hat g^{\mu\nu}\, ,\qquad \hat
g^{\mu\nu}\equiv-\frac{2}{\left(h^--h^+\right)^2}\, \left(
\begin{array}{cc}
2 & h^++h^- \\
h^++h^- & 2h^-h^+
\end{array}
\right)\, . \eqno (A.5)
$$
Here $\sqrt{-g}=e^{2F}\sqrt{-\hat g}\equiv e^{2F}\frac{h^--h^+}{2}$.

In tangent Minkowski space we also introduce \textit{light cone} coordinates
$x^a$ $(a=+,-)$, with $x^\pm\equiv\frac 12\, (x^0\pm x^1)$, so that raising
and lowering of the tangent space indices are performed as $A_\pm=-2A^\mp$.

The Riemannian connection on $\Sigma$ is defined by
$$
\label{connection} \omega_a=\varepsilon^{bc}\,{e^\mu}_b\,\partial_c
e_{a\mu}\, , \eqno (A.6)
$$
where $\varepsilon^{mn}\,(\varepsilon^{01}=1)$ is the constant totally
antisymmetric tensor in the Minkowski space or, in \textit{light cone}
basis, $\varepsilon^{-+}=-\varepsilon^{+-}=\frac 12$. Written in terms of
variables $(A.2)$ and $(A.3)$, the connection becomes
$$
\label{conn.expli} \omega_\pm =e^{-(F\pm f)}\left[
\hat\omega_\pm\mp\hat\partial_\pm (F\mp f)\right],\qquad
\hat\omega_\pm\equiv\mp\frac{2\partial_1h^\mp}{h^--h^+}\,. \eqno (A.7)
$$

Covariant derivative acting on a field with $n$ spinor indices has to be $%
\triangledown_a=\partial_a+\frac n2\,\omega_a$, or
$$
\label{Scov} \triangledown_\pm=e^{(\pm n-1)F-(n\pm 1)f}\,
\hat\triangledown_\pm\, e^{\mp nF+nf}, \eqno (A.8)
$$
where $\hat\triangledown_a= \hat\partial_a+\frac n2\,\hat\omega_a$.

Dirac matrices, defined in tangent Minkowski space, satisfy the Clifford
algebra
$$
\label{clifford} \{ \gamma^m ,\gamma^n\} = 2\eta^{mn}\, . \eqno (A.9)
$$
We use a representation
$$
\label{gamma} \gamma^0= \left(
\begin{array}{cc}
0 & 1 \\
-1 & 0
\end{array}
\right)\, ,\quad \gamma^1= \left(
\begin{array}{cc}
0 & 1 \\
1 & 0
\end{array}
\right)\, ,\quad \gamma_5\equiv \gamma^0\gamma^1= \left(
\begin{array}{cc}
1 & 0 \\
0 & -1
\end{array}
\right)\, , \eqno (A.10)
$$
where we have the identity $\mathop{\rm Tr}\nolimits\left(
\gamma^m\gamma^n\gamma_5\right) = 2\varepsilon^{mn}$. Also for projective $%
\gamma$-matrices $\gamma^\pm=\frac 12\, (\gamma^0\pm\gamma^1)$ we have:
$$
\label{pm.gamma} \gamma^+= \left(
\begin{array}{cc}
0 & 1 \\
0 & 0
\end{array}
\right)\, ,\qquad \gamma^-= \left(
\begin{array}{cc}
0 & 0 \\
-1 & 0
\end{array}
\right)\, . \eqno (A.11)
$$

Majorana spinor is a Dirac spinor $\theta_\alpha\equiv \left(
\begin{array}{c}
\theta_+ \\
\theta_-
\end{array}
\right)$ satisfying Majorana condition $\theta=C\bar\theta^T$, where $%
\bar\theta\equiv \theta ^\dagger \gamma ^0$, and $C$ is the charge
conjugation matrix ($C^{-1}\gamma _\mu C=-\gamma _\mu ^T$). In the
representation $(A.10)$ and with $C=\gamma^0$, the Majorana spinors are
real, $\theta_\alpha^*=\theta_\alpha$. Tensor $C^{\alpha\beta}$ performs the
raising of spinor indices ($\theta^\alpha=C^{\alpha\beta}\theta_\beta$),
while $C_{\alpha\beta}$ performs their lowering ($\theta_\alpha=C_{\alpha%
\beta}\theta^\beta$). In the components, it gives $\theta_\pm=\pm\theta^\mp$%
. Spinor contraction is denoted by $\theta\xi\equiv\theta^\alpha\xi_\alpha=-%
\theta_\alpha\xi^\alpha$.

The spinor covariant derivative is
$$
\label{Sderiv} D_\alpha=\bar\partial_\alpha+i\left( \gamma^m\theta\right)
_\alpha \partial_m \, , \eqno (A.12)
$$
where $\partial_m\equiv\frac{\partial}{\partial x^m}$ and $%
\bar\partial_\alpha\equiv\frac{\partial}{\partial\bar\theta^\alpha}$. More
explicitly, the derivative $(A.12)$ in the representation $(A.10)$ is
$$
\label{Dexpli} D^\pm\equiv\partial_{\theta_\mp}-i\theta_\mp\partial_\pm\, , %
\eqno (A.13)
$$
where $D_\alpha\equiv\left(
\begin{array}{c}
-D^+ \\
D^-
\end{array}
\right)$.

A generalization of the $\delta$-function to the super $\delta$-function is
$$
\label{Sdelta} \delta_{\pm 12}\equiv\theta_{\mp
12}\,\delta(x_1^\pm-x_2^\pm)\, , \eqno (A.14)
$$
where $\theta_{12}=\theta_1-\theta_2$. Its properties are
\[
\int d^4z_1\delta_{\pm 12}=1
\]
\[
F(z_1)\delta_{\pm 12}=F(z_2)\delta_{\pm 12}
\]
$$
\delta_{\pm 21}=-\delta_{\pm 12}\, ,\quad D^\pm_1\delta_{\pm 12}=
-D^\pm_2\delta_{\pm 12}\, , \eqno (A.15)
$$
where $d^4z\equiv d^2xd^2\theta$ and basic integrals for Grassman odd
numbers are $\int d\theta =0$ and $\int d\theta\,\theta=1$. Four real
coordinates $z^A=(x^\pm,\theta_\pm)$, with $x^\pm$ \textit{light cone}
coordinates and $\theta_\alpha$ a Majorana spinor, parametrize N=1
superspace.

\section*{Acknowledgments}

This work was supported in parts by the Serbian Ministry of Science,
Technology and Developments, and Chilean FONDECYT grant 2010017. O. M.
thanks to MECESUP for a scholarship. The institutional support of a group of
Chilean companies (CODELCO, Dimacoffi, Empresas CMPC, MASISA S. A. and
Telef\'{o}nica del Sur) and Yugoslav ``Vin\v ca'' Institute of Nuclear
Sciences is also acknowledged. CECS is a Millennium Science Institute.


\end{document}